# Co-doping effects on magnetism and superconductivity in the 112-type EuFeAs$_2$ system

Jia Yu[1,4*], Tong Liu[1,2*], BinBin Ruan[1,3], Kang Zhao[1,2], QingSong Yang[1,2], MengHu Zhou[1,3], ZhiAn Ren[1,2,3†]

[1]*Institute of Physics and Beijing National Laboratory for Condensed Matter Physics, Chinese Academy of Sciences, Beijing 100190, China*
[2]*School of Physical Sciences, University of Chinese Academy of Sciences, Beijing 100049, China*
[3] *Songshan Lake Materials Laboratory, Dongguan, Guangdong 523808, China*
[4]*School of Physics, Sun Yat-sen University, Guangzhou, Guangdong 510275, China*

The discovery of EuFeAs$_2$, currently the only charge-neutral parent phase of the 112-type iron-pnictide system, provides a new platform for the study of elemental doping effects on magnetism and superconductivity (SC). In this study, a series of polycrystalline EuFe$_{1-y}$Co$_y$As$_2$ and Eu$_{0.9}$Pr$_{0.1}$Fe$_{1-y}$Co$_y$As$_2$ samples are synthesized through solid-state reaction, and the evolutions of SC and magnetism with Co doping in EuFeAs$_2$ and Eu$_{0.9}$Pr$_{0.1}$FeAs$_2$ are investigated by electrical transport and magnetic susceptibility measurements. For EuFe$_{1-y}$Co$_y$As$_2$, the Eu-related antiferromagnetic (AFM) transition around 40 K is barely affected by Co doping, while the Fe-related spin density wave (SDW) transition temperature drops rapidly. Meanwhile, SC is induced by a trace amount of Co doping, with a highest transition temperature $T_c$ ~ 28 K found in EuFe$_{0.9}$Co$_{0.1}$As$_2$. For the Eu$_{0.9}$Pr$_{0.1}$Fe$_{1-y}$Co$_y$As$_2$ series, the magnetism and superconductivity show similar evolutions upon Co doping, and the highest $T_c$ is enhanced to 30.6 K with an optimum doping level $y$ ~ 0.07. Our results shed light on the competition between SC and SDW with Co doping in the 112-type EuFeAs$_2$ system.

**Keywords:** EuFeAs$_2$, Iron-based superconductors, Superconductivity, Magnetism
**PACS:** 74.70.–b, 74.70.Xa, 74.25.–q, 75.50.Ee

## 1. Introduction

The relationship between superconductivity (SC) and magnetism plays a critical role in understanding the pairing mechanism of high-temperature superconductivity [1, 2]. The iron-based superconductors form a big family of high-temperature SC, whose parent compounds are usually bad metals with Fe-related spin density wave (SDW) orders formed at low temperatures [3, 4]. Usually, SC can be induced from the parent phase either through chemical doping or by applying external pressure, with the SDW order suppressed simultaneously [5, 6]. The iron-based superconducting family manifests rich phase diagrams upon chemical doping. For instance, in the 1111-type CeFeAs(O,F) [4] or LaFeAs(O,F) [7], the SDW transition temperature decreases with increasing the F content, and SC emerges as soon as the SDW order is suppressed. In LaFeAs(O,H), the SDW order is gradually destroyed by doping in the underdoped region, whereas another SDW phase appears in the heavily (~ 40-50%) doped region [8]. Between these two SDW phases, a hump-like SC phase emerges with H doping, which leads to a unique bipartite phase diagram for the LaFeAsO system. Likewise, a second SC/SDW phase emerges individually in LaFeAs(O,F)/SmFeAs(O,H) in the heavily doped region [9, 10]. For the 122-type $A$Fe$_2$As$_2$ ($A$ = Alkaline earth metals or Eu) system, there is often a coexisting region of SC and SDW in the doping phase diagram [11-13]. Despite the variety of doping phase diagrams for iron pnictides, there is always an intense suppression effect of charge doping on SDW to induce SC. The same is true when taking the heavily doped LaFeAs(O,H) or SmFeAs(O,H) as a secondary parent phase.

---

[*] L. T. and Y. J. contributed equally to this work.
[†] Corresponding author. E-mail: renzhian@aphy.iphy.ac.cn

The 112-type CaFeAs$_2$ system [14-17] exhibits an unusual relationship between SDW and SC. The SDW is enhanced by La doping in (Ca,La)FeAs$_2$ and exists in all the experimentally obtainable overdoped samples [18, 19]. On the other hand, codoping with a small amount of transition metal suppresses the robust SDW order and enhances SC in (Ca,RE)FeAs$_2$ (RE = Rare earth elements) [20-24]. The suppression of the SDW order favors the inducing of SC as in most iron-pnictide systems. Besides, the existence of the As-chain layers makes Co doping more effective in inducing SC, because the La-doping-introduced extra electrons are believed to be partially distributed to the metallic As-chain layers [19, 25, 26]. Furthermore, Co codoping in the non-superconducting La-overdoped Ca$_{0.74}$La$_{0.26}$FeAs$_2$ can not only suppress the SDW order effectively, but also induce SC in this electron-overdoped system [27].

Recently, another 112-type iron pnictide EuFeAs$_2$ was successfully synthesized [28]. Two magnetic phase transitions were observed in single crystalline EuFeAs$_2$: the Eu-related antiferromagnetic (AFM) transition at 46 K, and the Fe-related SDW transition around 98 K [28, 29]. The SDW transition was proposed to follow a structural transition around 110 K. The SDW order can be slightly suppressed by 15% La doping, showing similar robustness as in the Ca112 system. Meanwhile, SC is induced with a highest $T_c \sim 11$ K. The lower $T_c$ in (Eu,La)FeAs$_2$ than those in many other iron pnictides [30, 31] is probably due to the influence of the strong magnetism of Eu$^{2+}$ as in (Eu,La)Fe$_2$As$_2$ [32, 33] and the competition from the robust SDW. Transition metal doping on the Fe site has been realized in Eu(Fe,Ni)As$_2$ [34, 35]. The SDW order is greatly suppressed by merely 4% Ni doping, and SC is enhanced with a higher $T_c \sim 17.5$ K. Given the competing relationship between SDW and SC in most of the iron pnictide systems, as well as the nature that Co doping suppresses the SDW order effectively in the Ca112 system, there is a promising possibility that Co doping in EuFeAs$_2$ will also suppress the SDW order significantly and enhance SC with a much higher $T_c$.

In this report, Co-doping effects on magnetism and SC in the EuFeAs$_2$ system are studied based on a series of polycrystalline EuFe$_{1-y}$Co$_y$As$_2$ and Eu$_{0.9}$Pr$_{0.1}$Fe$_{1-y}$Co$_y$As$_2$ samples. The SDW orders for both the basal compounds (EuFeAs$_2$ and Eu$_{0.9}$Pr$_{0.1}$FeAs$_2$) are indeed quickly suppressed through Co doping, and SC is induced/enhanced with much higher $T_c$s than those of the SC-SDW coexisting Eu$_{1-x}$La$_x$FeAs$_2$ system. Meanwhile, the Eu-related magnetism is barely affected by the substitution of Co for Fe, unlike the La-doping effect in Eu$_{1-x}$La$_x$FeAs$_2$.

## 2. Materials and methods

The polycrystalline EuFe$_{1-y}$Co$_y$As$_2$ ($y$ = 0, 0.03, 0.05, 0.08, 0.1, 0.12, 0.15) and Eu$_{0.9}$Pr$_{0.1}$Fe$_{1-y}$Co$_y$As$_2$ ($y$ = 0, 0.01, 0.02, 0.03, 0.05, 0.07, 0.08, 0.1, 0.12, 0.15, 0.20) samples were synthesized using the solid-state reaction method. First, EuAs, PrAs, FeAs, and CoAs precursors were prepared by the reactions of stoichiometric metals with arsenic powder at 1123-1173 K for two days. Second, the powdered precursors were mixed according to the stoichiometric ratio of EuFe$_{1-y}$Co$_y$As$_2$ or Eu$_{0.9}$Pr$_{0.1}$Fe$_{1-y}$Co$_y$As$_2$, ground thoroughly and pressed into pellets. The pellets were placed in alumina crucibles, sealed into argon filled quartz tubes, then heated slowly to 1073 K and held for 40 hours. Since EuAs can be easily oxidized, all preparing manipulations were carried out in a glove box protected with high-purity argon gas.

The crystal structure was identified by powder X-ray diffraction (PXRD) performed on a PANalytical's powder X-ray diffractometer. The Energy-Dispersive X-ray Spectroscopy (EDS) investigation was performed on a Phenom's scanning electron microscope (SEM). The resistivity was measured by the standard four-probe method using a Quantum Design physical property measurement system (PPMS). The magnetic susceptibility was measured by a Quantum Design magnetic property measurement system (MPMS).

## 3. Results and discussion

PXRD patterns for the EuFe$_{1-y}$Co$_y$As$_2$ and Eu$_{0.9}$Pr$_{0.1}$Fe$_{1-y}$Co$_y$As$_2$ samples are shown in Figure 1(a) and (b),

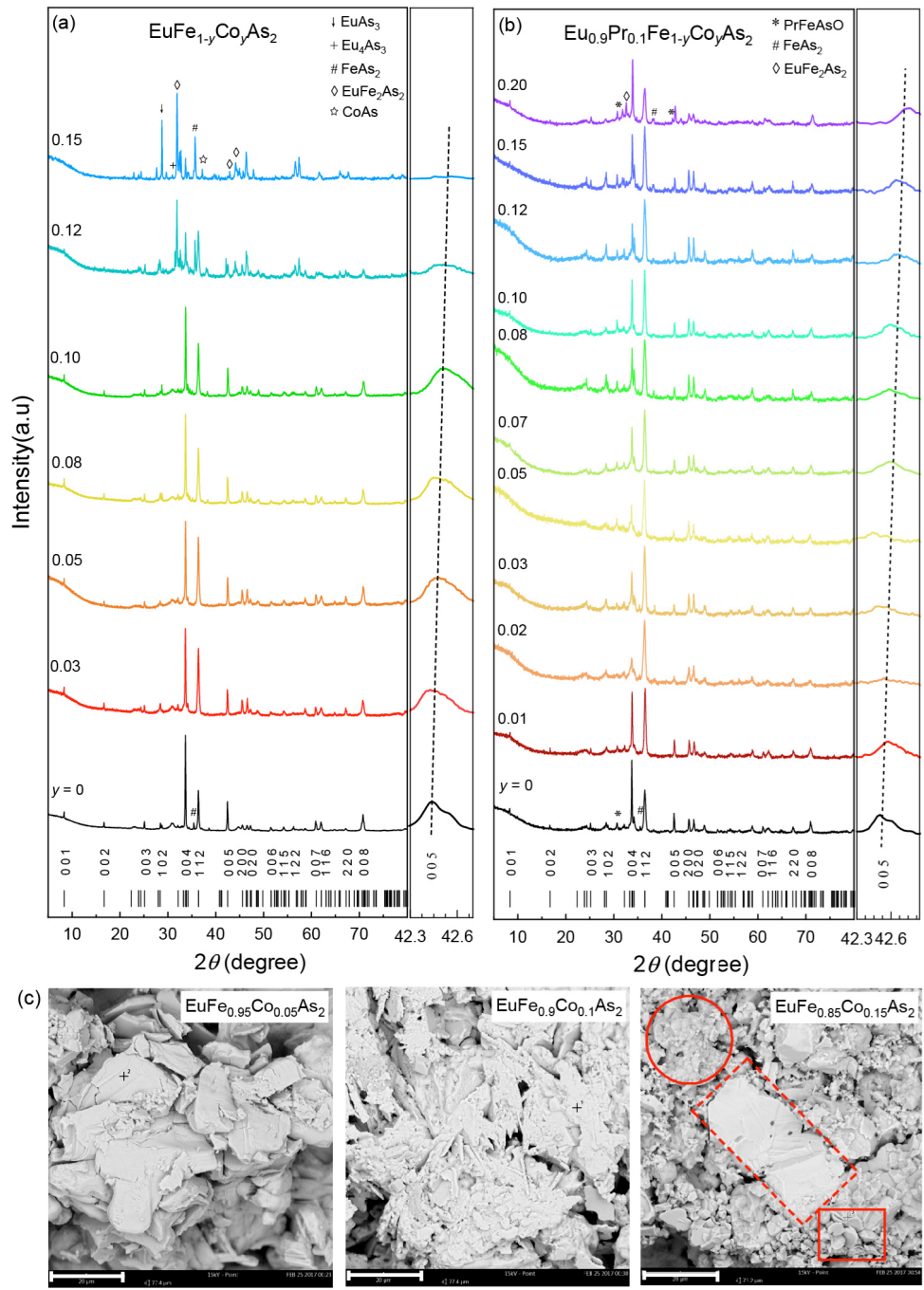

Figure 1 PXRD patterns for (a) EuFe$_{1-y}$Co$_y$As$_2$ and (b) Eu$_{0.9}$Pr$_{0.1}$Fe$_{1-y}$Co$_y$As$_2$. The bars at the bottom denote the calculated Bragg positions of EuFeAs$_2$ under the space group $P2_1/m$. All the Miller indexes are based on the space group $P2_1/m$. Enlarged views of the diffraction patterns around (0 0 5) are attached to the right. (c) SEM images for EuFe$_{1-y}$Co$_y$As$_2$ ($y$ = 0.05, 0.1, 0.15), where the crystals in the circle, the dashed box, and the solid box in the right panel mainly belong to the Eu112 phase, the Eu-As binary phase, and the Eu122 phase, respectively. The length of the scale marks is 20 μm.

respectively. According to the previously reported single crystal X-ray diffraction results, La doping in EuFeAs$_2$ induces a structural transformation, from orthorhombic $I/mm2$ to monoclinic $P2_1/m$ [28, 29]. However, the structural configuration of the alternation between the FeAs-layers and the As-planes remains generally the same as in (Ca,$RE$)FeAs$_2$ for both structures. Polycrystalline samples of the EuFeAs$_2$ system show poor crystallinity due to the narrow reaction window [28, 34]. Fortunately, the calculated diffraction peaks of the two space groups share almost the same $2\theta$ positions. Thus, the observed main peaks could be indexed under either of the space groups. Here, all the PXRD patterns for the Co-doped and the Pr/Co-codoped samples in this article are indexed under the space group $P2_1/m$ for consistency with the previous reports [28, 34]. The $2\theta$ positions of the observed diffraction peaks for EuFeAs$_2$ and Eu$_{0.9}$Pr$_{0.1}$FeAs$_2$ match well with the calculated ones, except for a few low-intensity peaks associated with trace impurities. Upon Co doping, the (0 0 5) peak shifts slightly to higher degrees for both systems, indicating a shrinkage of the lattice in the $c$ direction. For EuFe$_{1-y}$Co$_y$As$_2$, the EuFe$_2$As$_2$ and CoAs impurities emerge and become significant when the Co content $y$ exceeds 0.1, which suggests that the solubility of Co in EuFeAs$_2$ is about 10%. Comparatively, the solubility of Co in Eu$_{0.9}$Pr$_{0.1}$FeAs$_2$ is much higher according to the PXRD patterns, where only minor impurities are observed when the Co-doping level is as high as 0.2. The enhanced structural stability by introducing Pr doping, somehow, is in agreement with the rare earth metal doping effect in the undoped-phase-absent (Ca,RE)FeAs$_2$ system.

To further study the crystallinity and the chemical phases, we performed an EDS investigation on the Co-doped EuFe$_{1-y}$Co$_y$As$_2$ ($y$ = 0.05, 0.1, 0.15) samples. Typical SEM images of the as-grown pellets' surfaces are demonstrated in Figure 1(c). For $y$ = 0.05, the sample consists of small plate-like grains with uniform appearance, determined to be the Eu112 phase by EDS analysis in different areas. The grains exhibit curved edges with a characteristic appearance of an alloy after melting, which we ascribe to the relatively low melting point of EuAs$_2$ (~ 1073 K) serving as the flux during the synthetic process. The plate-like and curve-edged shapes of the grains are responsible for the high intensity of the XRD peaks in the preferred orientation and the weak diffraction in other orientations. For $y$ = 0.1, the sample also mainly consists of plate-like Eu112 grains with curved edges. With the Co-doping level increasing, the general sizes of the grains decrease, and the specific area of the curved edges increases. For $y$ = 0.15, the sample consists of a small amount of Eu112 phase and numerous impurities. The grains of the Eu112 phase, see the area in the solid circle, degenerate into small ingots with curved surfaces. The EDS analysis agrees with the PXRD results that the samples with the Co-doping level $y \leq 0.1$ mainly consist of the Eu112 phase, while, the impurities become major components with further doping.

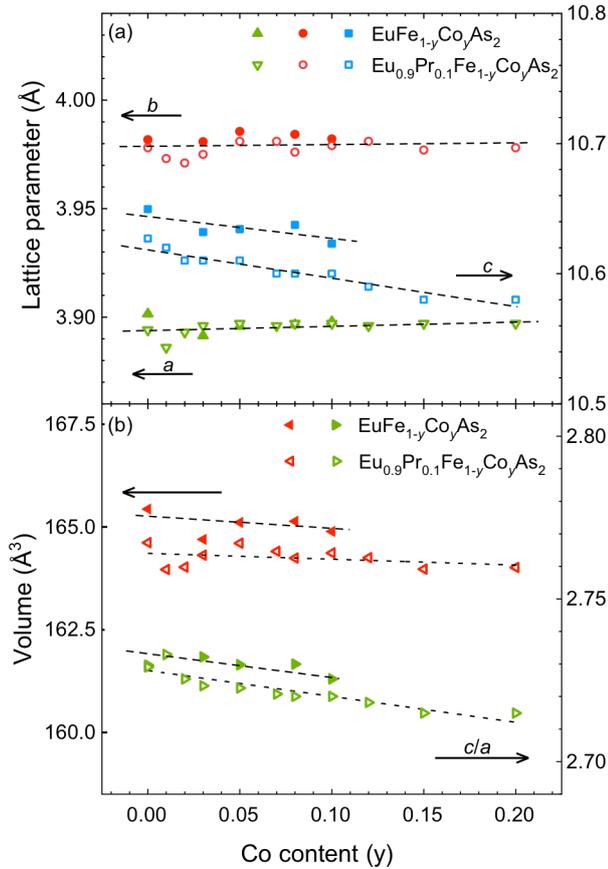

Figure 2 The evolutions of (a) lattice parameters, and (b) cell volume and $c/a$ value, for EuFe$_{1-y}$Co$_y$As$_2$ and Eu$_{0.9}$Pr$_{0.1}$Fe$_{1-y}$Co$_y$As$_2$ with Co doping.

The calculated lattice parameters for EuFe$_{1-y}$Co$_y$As$_2$ and Eu$_{0.9}$Pr$_{0.1}$Fe$_{1-y}$Co$_y$As$_2$ are summarized in Figure 2(a). Slight lattice shrinkage in the $c$ direction occurs for both systems. Meanwhile, no significant change is observed in the other two directions. As a result, the cell volume also shrinks slightly upon Co doping, as depicted in Figure 2(b), which agrees with the smaller ionic radius of $Co^{2+}$ than that of $Fe^{2+}$ and indicates a monotonic increase in the actual Co-doping concentration.

To investigate the Co-doping effects on the magnetism and SC, we performed electrical transport measurements on the Co-doped and Pr/Co-codoped compounds, as included in Figure 3. All the resistivity curves show linear metallic behavior at high temperatures. For $EuFe_{1-y}Co_yAs_2$, as shown in Figure 3(a), the anomaly related to the structural/SDW transitions around 100 K in the parent phase is quickly suppressed to about 83 K by merely 3% Co doping, with SC induced simultaneously. The single peak in the $dR/dT$ curve for $y = 0.03$ implies that the structural and SDW transitions might occur simultaneously. Co doping seems to converge the two transitions together, rather than to diverge them as in the 122-type $Eu(Fe_{1-x}Co_x)_2As_2$ [36]. No sign of the structural and SDW transitions appears above 50 K when the doping level reaches 5%. The anomaly related to the Eu-AFM transition around 40 K (indicated by the red arrow) is barely affected by Co doping, implying that the Eu-AFM order is robust against Co doping.

The optimum Co-doping level for $EuFe_{1-y}Co_yAs_2$ is $y = 0.1$, with SC observed at $T_c \sim 28$ K, as shown in the lower inset in Figure 3(a). The much higher $T_c$ of $EuFe_{1-y}Co_yAs_2$ than that of the optimum doped $(Eu,La)FeAs_2$ suggests that the suppression of the SDW order indeed improves SC in the EuFeAs$_2$ system. Moreover, to our knowledge, 28 K is the highest $T_c$ of all the merely-Co-doped iron pnictides ever reported, which may also be associated with the suppression of the robust SDW. The novel Ca112 system with the coexistence of SDW and SC shows a highest $T_c \sim 47$ K [16], implying a strong superconducting pairing via the unusually robust SDW in the 112-system. Here, when the robust SDW order is suppressed through Co doping in EuFe$_{1-y}$Co$_y$As$_2$, the corresponding spin fluctuation may still lead to a strong pairing, thus resulting in a higher $T_c$ than in other Co-doped iron pnictides. The temperature dependences of resistivity ($R$-$T$) for the optimum doped EuFe$_{0.9}$Co$_{0.1}$As$_2$ were measured under different magnetic fields from 0 to 9 T, as depicted in Figure 3(b). The relation between the mid-point of the superconducting transition ($T_c^{mid}$) and the applied magnetic field is roughly fitted with the Ginzburg-Landau (G-L) model:

$$H_{c2}(T) = H_{c2}(0)[(1-t^2)/(1+t^2)], \qquad (1)$$

where $t$ represents the normalized temperature $T/T_c^{mid}$.

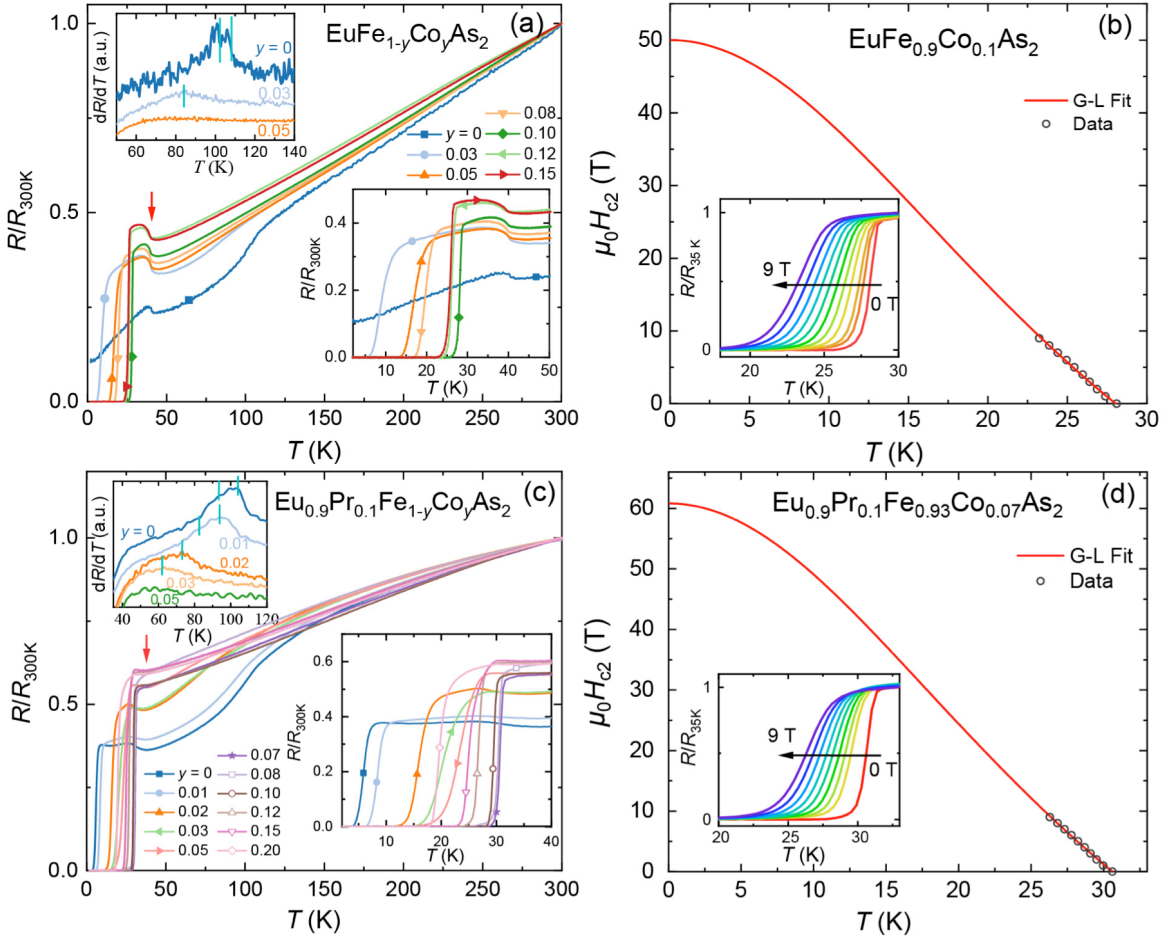

Figure 3 The normalized resistivity versus temperature for (a) $EuFe_{1-y}Co_yAs_2$, (c) $Eu_{0.9}Pr_{0.1}Fe_{1-y}Co_yAs_2$ under zero magnetic field. Upper insets in (a) and (c) show the first derivation of resistivity, and the lower insets the enlarged demonstration of the superconducting transitions. The G-L fitting curves of $\mu_0H_{c2}$-$T$ of for (b) $EuFe_{0.9}Co_{0.1}As_2$ and (c) $Eu_{0.9}Pr_{0.1}Fe_{0.93}Co_{0.07}As_2$, with the resistivity under different magnetic fields displayed in the insets. The resistivity data for the undoped $EuFeAs_2$ in (a) are extracted from our previous work [28].

For the Pr/Co-codoping Eu$_{0.9}$Pr$_{0.1}$Fe$_{1-y}$Co$_y$As$_2$ series, as demonstrated in Figure 3(c), the basal Eu$_{0.9}$Pr$_{0.1}$FeAs$_2$ shows similar resistivity anomalies related to the AFM, SDW, and structural transitions around 38 K, 93 K, and 104 K, respectively. The structural and SDW transitions are also suppressed fast by Co doping and become indistinguishable with the doping level exceeding 0.03, as shown in the upper inset of Figure 3(c). The Eu-AFM transition is also insensitive to Co doping as seen from the kinks indicated by the red arrow (not clear for the samples with the doping levels around 0.08). Because the basal Eu$_{0.9}$Pr$_{0.1}$FeAs$_2$ is electron doped, the optimum doping level $y = 0.07$ is reasonably lower than that for EuFe$_{1-y}$Co$_y$As$_2$, as seen from the enlarged view around the superconducting transitions in the lower inset of Figure 3(c). In the meantime, the highest $T_c$ is enhanced to 30.6 K in this Pr-tuned system. Figure 3(d) shows the $R$-$T$ curves under different fields and the G-L fitted $\mu_0 H_{c2}$-$T$ curve for the optimum doped Eu$_{0.9}$Pr$_{0.1}$Fe$_{0.93}$Co$_{0.07}$As$_2$.

To further investigate the Eu-related magnetism and the superconducting diamagnetism, the temperature dependences of the DC magnetic susceptibility ($\chi$-$T$) for both systems were measured by the zero-field-cooling (ZFC) and field-cooling (FC) methods, as demonstrated in Figure 4. For the EuFe$_{1-y}$Co$_y$As$_2$ series, the data of the samples containing the 122-phase ($y = 0.12$ and 0.15) impurities are not presented owing to the dramatic ferromagnetic (FM) response from the canted Eu$^{2+}$ moment of Eu(Fe$_{1-x}$Co$_x$)$_2$As$_2$ [36].

The Eu-AFM phase transitions occur around $T_N \sim 41$ K for all the EuFe$_{1-y}$Co$_y$As$_2$ samples, as shown in Figure 4(a), which agrees with the results of the resistivity measurements. A slight bifurcation between the ZFC and FC curves below $T_N$ appears with 3% Co doping, and remains along with the doping level increasing. This slight bifurcation implies a mild moment canting of Eu$^{2+}$. The mild canting nature indicates that the Eu-related AFM order is less sensitive to Co doping in EuFe$_{1-y}$Co$_y$As$_2$ than in the 122-type Eu(Fe$_{1-x}$Co$_x$)$_2$As$_2$ with doping-induced AFM-to-FM transformation [36]. The robustness of the Eu-related AFM order upon doping in EuFe$_{1-y}$Co$_y$As$_2$ is probably due to the presence of the As-planes. In the 122-type Eu(Fe$_{1-x}$Co$_x$)$_2$As$_2$, the magnetic ground state is modifiable by tuning the 3$d$-electrons-mediated Ruderman-Kittel-Kasuya-Yosida (RKKY) interaction between the Eu$^{2+}$ moments, as well as the direct interactions between the Fe$^{2+}$ and Eu$^{2+}$ moments through Co doping [36]. In the 122-type structure, each Eu-layer is sandwiched by two adjacent FeAs-layers. However, in the 112-type structure, each Eu-layer is sandwiched by one FeAs-layer and an As-chain layer. Compared with Eu(Fe$_{1-x}$Co$_x$)$_2$As$_2$, the extra intermediate $d$-electrons from Co doping are halved in EuFe$_{1-y}$Co$_y$As$_2$ for each Eu$^{2+}$ ion, at the same doping level. Meanwhile, there are eight nearest Fe$^{2+}$ ions for each Eu$^{2+}$ ion to provide the direct interaction between the Fe$^{2+}$ and Eu$^{2+}$ moments in the 122-type structure, which are also halved in the 112-type structure. As a result, the Eu-related magnetism is less sensitive to Co doping in EuFe$_{1-y}$Co$_y$As$_2$ than in Eu(Fe$_{1-x}$Co$_x$)$_2$As$_2$.

With temperature going down, a superconducting diamagnetic behavior appears for EuFe$_{1-y}$Co$_y$As$_2$ with trace doping. The diamagnetism is insignificant with the Co doping level $y = 0.03$, but greatly enhanced through further doping, see the $\chi$-$T$ curves zoomed out in the inset of Figure 4(a). The values of the shielding volume fraction (VF) at 2 K are over 40% for $y = 0.05$, 0.08, and 0.1, indicating the emergence of bulk SC in the Co-doped EuFe$_{1-y}$Co$_y$As$_2$.

A remarkable increase of the magnetization appears around 15 K in the FC curves, which is in agreement with the spin-glass-like (SG-like) transition observed in polycrystalline EuFeAs$_2$ under a lower magnetic field of 10 Oe [34]. Besides, a spin-reorientation-like (SR-like) behavior occurs with a slight upturn around 7 K. The SR-like transition probably originates from the magnetic anisotropic interaction(s) [37]: the anisotropic-symmetric interaction between Eu$^{2+}$ and Fe$^{2+}$, and/or the possible antisymmetric Dzyaloshinsky–Moria (DM) interaction between the Eu$^{2+}$ moments in the low-symmetric structure. Meanwhile, the SDW transition is indistinguishable from the Eu-related magnetic background, and the magnetic susceptibility for higher temperatures exhibits general Eu-related paramagnetism (not demonstrated).

For Eu$_{0.9}$Pr$_{0.1}$Fe$_{1-y}$Co$_y$As$_2$, as shown in Figure 4(b) and the upper inset, diamagnetic transitions are realized for all the Co-codoped samples under a low magnetic field of 10 Oe, with a highest transition temperature around 30 K for $y = 0.07$. The enhanced VF indicates an improvement of SC by Co-coping from the basal Eu$_{0.9}$Pr$_{0.1}$FeAs$_2$. The SG-like behavior around 15 K remains, while the SR-like upturn disappears with Pr/Co codoping. Slight moment canting also exists in the Pr/Co-codped Eu$_{0.9}$Pr$_{0.1}$Fe$_{1-y}$Co$_y$As$_2$. The bifurcations of the $\chi$-$T$ curves occur above the $T_N$s, which indicates a Griffiths-like phase induced by Pr/Co codoping. The moment canting and the Griffiths-like behavior suggest a change of the competitive balance between the FM and AFM interactions, which leads to the bump-like AFM transition, corresponding to the broaden anomalies in the $R$-$T$ curves. Similar to the electrical transport results, the Eu-AFM transition is unidentifiable for the samples with the doping levels around 0.08. To further identify the magnetic transition temperatures, FC magnetic susceptibility measurements are performed under a larger magnetic field of 1 T for the Eu$_{0.9}$Pr$_{0.1}$Fe$_{1-y}$Co$_y$As$_2$ samples with $y \leq 0.12$, as demonstrated in the lower inset in Figure 4(b). The Eu-related AFM transition is determined to occur around 30 K for $y \sim 0.8$, coincidentally overlapping with the superconducting transition observed in the resistivity and low-field susceptibility measurements.

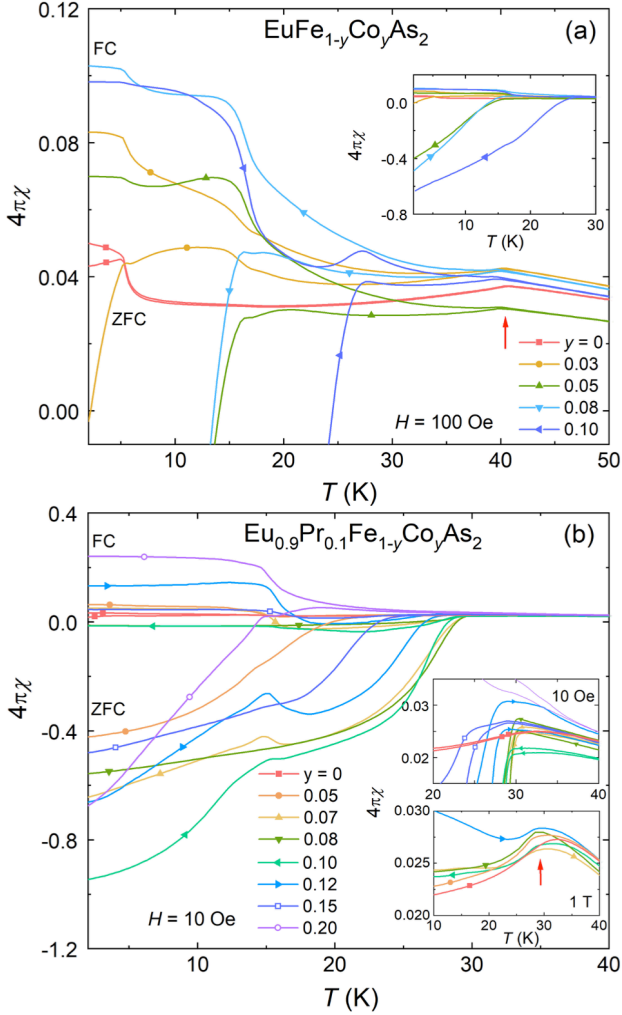

Figure 4 The temperature dependence of magnetic susceptibility for (a) EuFe$_{1-y}$Co$_y$As$_2$ under a magnetic field of 100 Oe and (b) Eu$_{0.9}$Pr$_{0.1}$Fe$_{1-y}$Co$_y$As$_2$ under a magnetic field of 10 Oe, for both ZFC and FC measurements. The paramagnetic parts for higher temperatures are not displayed. The inset in (a) shows the zoomed out $\chi$-$T$ curves. The upper and lower insets in (b) show the enlarged details of the superconducting transition ($H = 10$ Oe) and Eu-related AFM transition ($H = 1$ T), respectively. The susceptibility data for the undoped EuFeAs$_2$ in (a) are extracted from our previous work [28].

## Summary

The Co-doping effects on the SDW, Eu-related magnetism and superconductivity in the Eu112-type $EuFe_{1-y}Co_yAs_2$ and $Eu_{0.9}Pr_{0.1}Fe_{1-y}Co_yAs_2$ are studied systematically through PXRD, EDS, electrical transport, and magnetic susceptibility measurements. The electrical and magnetic results are summarized in a phase diagram in Figure 5, including the evolutions of the SDW, the Eu-related magnetic, and the superconducting transition temperatures as functions of the Co-doping level $y$ for both systems. Compared with the La-doped $(Eu,La)FeAs_2$ system, Co doping suppresses the SDW transition more effectively in both the undoped $EuFeAs_2$ and the Pr-tuned $Eu_{0.9}Pr_{0.1}FeAs_2$. As a result, SC with a much higher $T_c$ than that in the optimum doped SDW-SC-coexisting $(Eu,La)FeAs_2$ is induced, which is a reflection of the SDW-SC competing nature in the Eu112 system. The optimum doped $EuFe_{0.9}Co_{0.1}As_2$ exhibits SC with $T_c \sim 28$ K. The greater efficiency of Co doping than La doping may also be associated with the existence of the As-chain layers, as in the Ca112 system. By Pr-doping tuning, the optimum SC can be further enhanced to $T_c \sim 30.6$ K. On the other hand, the Eu-related AFM order is insensitive to Co doping in both systems, but still presents a mild moment canting.

The Co-doping effects on suppressing the Fe-related SDW and improving SC make Co doping a unique way to investigate the relationship between SC and SDW in this novel system, and to search higher $T_c$ in the 112-type iron pnictides.

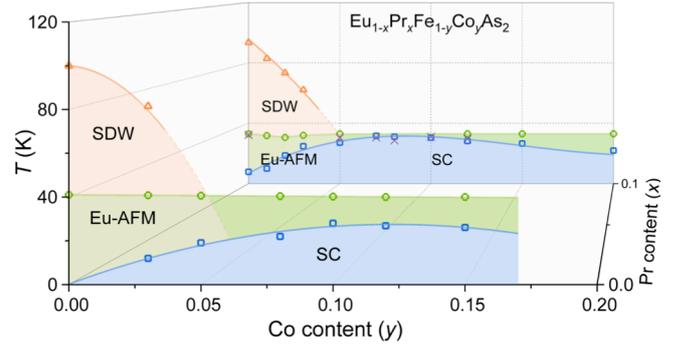

Figure 5 Combined phase diagram on the SDW order, the Eu-related magnetic order, and SC for $Eu_{1-x}Pr_xFe_{1-y}Co_yAs_2$ ($x = 0$ and 0.1). The triangles (squares) represent the SDW (superconducting) transition temperatures obtained from the $R$-$T$ curves. For $x = 0$, the circles represent the Eu-related transition temperatures obtained from the $\chi$-$T$ curves. For $x = 0.1$, the circles and the crosses represent the Eu-related transition temperatures obtained from the corresponding anomalies (mid-points) in the $R$-$T$ curves and from the high-field $\chi$-$T$ curves, respectively. The values for the undoped $EuFeAs_2$ are extracted from our previous work [28].

## Acknowledgments

The authors are grateful for the financial support from the National Natural Science Foundation of China (11774402) and the National Key Research Program of China (2018YFA0704200, 2016YFA0300301).